\documentclass[10pt,twocolumn,letterpaper]{article}

\usepackage{iccv}              

\usepackage{graphicx}
\usepackage{amsmath}
\usepackage{amsfonts, amssymb}
\usepackage{booktabs}
\usepackage{tabularx}
\usepackage{multirow}
\usepackage[shortcuts]{extdash}

\definecolor{iccvblue}{rgb}{0.21,0.49,0.74}
\usepackage[pagebackref,breaklinks,colorlinks,allcolors=iccvblue]{hyperref}

\def\paperID{172} 
\def\confName{ICCV}
\def\confYear{2025}

\title{Neighboring Slice Noise2Noise: Self-Supervised Medical Image Denoising from Single Noisy Image Volume}

\author{Langrui Zhou$^1$, Ziteng Zhou$^1$, Xinyu Huang$^1$, Huiru Wang$^1$, Xiangyu Zhang$^1$, Guang Li$^{1*}$\\
$^1$School of Biological Science and Medical Engineering, Southeast University, Nanjing, China\\
{\tt\small liguang@seu.edu.cn}
}

\begin{document}
\maketitle

\begin{abstract}

\vspace{-1mm}

\indent In the last few years, with the rapid development of deep learning technologies, supervised methods based on convolutional neural networks have greatly enhanced the performance of medical image denoising. 
However, these methods require large quantities of noisy-clean image pairs for training, which greatly limits their practicality.
Although some researchers have attempted to train denoising networks using only single noisy images, existing self-supervised methods, including blind-spot-based and data-splitting-based methods, heavily rely on the assumption that noise is pixel-wise independent.
However, this assumption often does not hold in real-world medical images.
Therefore, in the field of medical imaging, there remains a lack of simple and practical denoising methods that can achieve high-quality denoising performance using only single noisy images.
In this paper, we propose a novel self-supervised medical image denoising method, Neighboring Slice Noise2Noise (NS-N2N).
The proposed method utilizes neighboring slices within a single noisy image volume to construct weighted training data, and then trains the denoising network using a self-supervised scheme with regional consistency loss and inter-slice continuity loss.
NS-N2N only requires a single noisy image volume obtained from one medical imaging procedure to achieve high-quality denoising of the image volume itself.
Extensive experiments demonstrate that the proposed method outperforms state-of-the-art self-supervised denoising methods in both denoising performance and processing efficiency.
Furthermore, since NS-N2N operates solely in the image domain, it is free from device-specific issues such as reconstruction geometry, making it easier to apply in various clinical practices.
Code is available at \href{https://github.com/LangruiZhou/Neighboring-Slice-Noise2Noise}{https://github.com/LangruiZhou/Neighboring-Slice-Noise2Noise}.
\end{abstract}

\vspace{-4mm}

\section{Introduction}
\indent \indent Advanced medical imaging technologies, such as computed tomography (CT), magnetic resonance imaging (MRI), and positron emission tomography (PET), play a crucial role in clinical disease diagnosis today.
However, the medical imaging process is often compromised by noise, leading to a decrease in image quality, which in turn negatively impacts subsequent disease diagnosis and clinical decision-making \cite{SS20}.
Therefore, denoising is of great importance in medical image processing.
Medical image denoising can be performed at different stages of imaging, including pre-processing, reconstruction optimization, and post-processing.
The former two often face device-specific issues, as they require specific geometric parameters to reconstruct the raw data into medical images.
In contrast, denoising methods applied directly to the reconstructed images can be more easily integrated into existing clinical procedures.
As a result, image-domain denoising methods have attained widespread attention in recent years and made considerable progress.

Traditional medical image denoising methods, such as NLM\cite{BA05} and BM3D\cite{DK07}, suppress noise to some extent by exploiting the correlations between similar regions in noisy images.
In recent years, with the rapid development of deep learning technologies, supervised denoising methods based on convolutional neural networks (CNNs) \cite{ZK17,MX16,RO15} have greatly improved image denoising performance and have been widely applied in the field of medical image denoising, including low-dose CT denoising\cite{CH17,YQ18}, MR image restoration under Rician noise\cite{JD18,YH22}, and low-dose PET denoising\cite{WY18,KK18}.
These methods require a large number of noisy-clean image pairs for training. However, in practical applications, acquiring a large set of well-aligned noisy-clean image pairs is expensive, time-consuming, and sometimes impossible.
Moreover, these supervised methods have limited generalization to unseen data and often require retraining to maintain good performance when encountering new data with significant distribution differences. These limitations greatly restrict the practicality of CNN-based supervised denoising methods.

To overcome these limitations, the classic Noise2Noise\cite{LJ18} method was proposed, which suggests that when the noise has zero mean, using another noisy image of the same scene as the training target leads to a denoiser with performance comparable to one trained with clean images.
Although clean images are no longer required during training, obtaining aligned noisy-noisy image pairs through just two scans is still impractical in medical settings. 
Therefore, a series of self-supervised methods that require only single noisy images have been proposed, primarily consisting of blind-spot-based methods and data-splitting-based methods.

Blind-spot-based methods, such as Noise2Void\cite{KA19}, Noise2Self\cite{BJ19}, and Self2Self\cite{QY20}, assume that the true values of neighboring pixels are highly correlated, while the noise is zero-mean and pixel-wise independent.
By constructing blind-spot structures to prevent the network from learning identity mapping, these methods encourage the network to use information from neighboring pixels to predict the true value of the current pixel, thereby achieving effective denoising.
Data-splitting-based methods, such as Neighbour2Neighbour\cite{HT21} and Zero-Shot Noise2Noise\cite{MY23}, also assume that the noise is zero-mean and pixel-wise independent.
By splitting single noisy images (\eg, through neighborhood down-sampling), these methods can generate a large number of paired independent noisy images, thereby extending the Noise2Noise training strategy to scenarios where only single noisy images are available. 
However, medical images do not always satisfy the assumption of pixel-wise independent noise.
The complex operations involved in medical image reconstruction, such as back-projection and filtering, often introduce noise correlations between neighboring pixels within each reconstructed slice image, which may negatively affect the performance of the self-supervised methods mentioned above.
Some researchers have attempted to apply the aforementioned self-supervised methods to the raw data before reconstruction\cite{LJ24, WD19} in order to avoid the issue of noise correlation in the image domain.
For example, Wu \etal\cite{WD21} obtained paired noisy images by dividing the CT projection data into odd and even subsets and reconstructing them separately, while Yu \etal\cite{YJ23} applied the Noise2Self\cite{BJ19} method to denoise low-dose CT projection data.
However, these methods need to consider device-specific issues, such as reconstruction geometry, which increases their application complexity and reduces usability.
In summary, there is still a lack of simple and practical denoising methods in the medical imaging field that can achieve high-quality denoising using only single noisy images. 

In fact, the existing self-supervised denoising methods can be viewed as adapting the Noise2Noise training strategy to scenarios with only single noisy images by artificially constructing noisy pixel pairs or noisy sub-image pairs. 
However, when dealing with medical images, there is a simpler and more efficient way to construct noisy image pairs.
Common medical images, such as CT, MR, and PET, are usually presented in the form of image volumes. 
Due to the spatial continuity of tissue structures, there are many regions belonging to the same tissue across neighboring slices within the same image volume, and these regions should theoretically have the same true values. 
For convenience, we refer to the regions between neighboring slices with the same 2D coordinates and the same true values as matched regions.
Since there is minimal interference between slices during the reconstruction process, the corresponding pixels within matched regions exhibit better noise independence.
Considering the high information redundancy in image data and the findings of previous research \cite{FC22, ZL24}, it is possible to adequately train the network and achieve good performance by applying the Noise2Noise training strategy only in the matched regions. 

In this work, we propose Neighboring Slice Noise2Noise (NS-N2N), a novel self-supervised medical image denoising method that requires only a single noisy image volume for training and can perform high-quality denoising of the image volume itself.
The proposed method features a training data construction approach that utilizes neighboring slice images to generate weighted image pairs, along with a self-supervised training scheme that incorporates regional consistency loss and inter-slice continuity loss.
To validate its performance, we conducted synthetic experiments using the IXI brain MRI dataset \cite{IXI18} and real-world experiments with the Mayo Low-dose CT Challenge dataset \cite{MC17}.
Both qualitative and quantitative results demonstrate that NS-N2N outperforms state-of-the-art self-supervised denoising methods that use only single noisy images for training, offering superior denoising performance and higher processing efficiency, especially on real-world datasets.

The main contributions of our paper are as follows:
\begin{enumerate}
    \item We propose a novel self-supervised method for medical image denoising that only requires a single noisy image volume from one medical imaging procedure for training, achieving high-quality denoising of the image volume itself.
    \item The proposed NS-N2N method is an image-domain approach that directly denoises the reconstructed images. Unlike denoising methods that process raw data, NS-N2N does not have device-specific issues, making it easier to apply in various clinical practices.
    \item Compared to state-of-the-art self-supervised denoising methods, the proposed NS-N2N method demonstrates notable superiority in denoising performance and processing efficiency, especially in real-world medical scenarios where the entire volume consisting of hundreds of noisy images needs to be denoised in a single pass.
\end{enumerate}

\section{Related Work}
\textbf{Learning-based Medical Image Denoising Methods}\\
\indent With the rapid advancement of deep learning, supervised denoising methods based on convolutional neural networks have achieved excellent performance in medical image denoising tasks.
For low-dose CT denoising, Chen \etal\cite{CH17} proposed a residual encoder-decoder convolutional neural network that learns the mapping between low-dose CT images and normal-dose CT images.
In contrast, Yang \etal\cite{YQ18} applied generative adversarial networks to this task, achieving even better performance.
As for MR, You \etal\cite{YX19} and Jiang \etal\cite{JD18} used convolutional neural networks to restore MR images degraded by Rician noise, while Yang \etal\cite{YH22} developed a hybrid residual MLP-CNN model for denoising MR image volumes.
For low-dose PET denoising, Sano \etal\cite{SA21} modified the U-Net network to directly generate full-dose PET images from low-dose PET images.
Additionally, several studies have employed generative adversarial networks to achieve low-dose PET denoising \cite{WY18, LW19, GY20}.
However, these methods require large quantities of noisy-clean image pairs for training, and often have limited generalization capability.

To overcome these limitations, some researchers have attempted to apply self-supervised image denoising frameworks to medical scenarios. 
Wu \etal\cite{WD21} and Liu \etal\cite{LJ24} obtained paired noisy images by splitting CT projection data into odd and even subsets and reconstructing them separately, where the Noise2Noise\cite{LJ18} training strategy could be applied.
Yu \etal\cite{YJ23} utilized Noise2Self\cite{BJ19} to directly denoise the CT projection data and then reconstructed the denoised projections to obtain clean images.
However, these methods all need some device-specific information, such as reconstruction geometry parameters, to reconstruct the raw data into medical images.
Therefore, these methods have restricted usability, as we often cannot obtain device-specific information from third-party devices.
There are also self-supervised methods that operate solely in the image domain.
For example, Deformed2Self\cite{XJ21} achieves self-supervised denoising of dynamic imaging results by applying spatial motion modeling to slices from different time frames.
However, this method is only applicable in scenarios where images from different time frames can be obtained, which greatly limits its range of applications.
In summary, the medical image field still lacks simple and practical image-domain denoising methods that can achieve high-quality denoising results using only single noisy images.

\section{Method}
\indent \indent The main idea of the proposed NS-N2N method is to construct weighted training data using neighboring slice images from a single noisy image volume and use this weighted data for self-supervised training of the network, thereby achieving high-quality denoising. 
We begin with a brief review of the classic Noise2Noise training scheme and the motivation behind our research.
Next, we explain how training data are constructed by generating weighted image pairs from neighboring slice images. 
Finally, we present the self-supervised training scheme with regional consistency loss and inter-slice continuity loss.

\begin{figure*}[htb]
    \centering
    \includegraphics[width=0.6\linewidth]{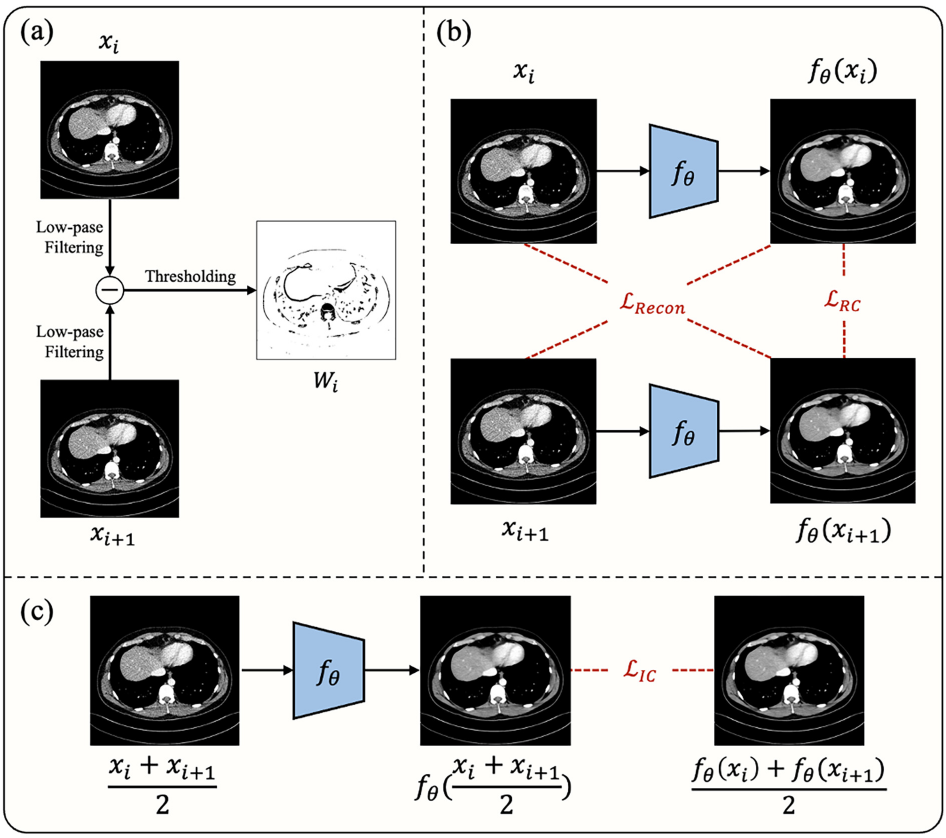}
    \caption{Overview of the proposed NS-N2N method. (a) The proposed weight matrix calculation method. (b)(c) The proposed self-supervised training scheme with regional consistency loss and inter-slice continuity loss.}
    \label{fig1}
    \vspace{-4mm}
\end{figure*}

\subsection{Motivation}
\indent \indent Noise2Noise\cite{LJ18} is a classic denoising method that can be trained without the need for clean images. 
Given an image pair consisting of two independent noisy observations of the same scene, $x_1=s+n_1$ and $x_2=s+n_2$, where $s$ represents the true values (clean image), and $n_1$ and $n_2$ are noise terms with zero mean, the process of training a denoising network $f_\theta$ to map one noisy image $x_1$ to the other noisy image $x_2$ is, in terms of expectation, equivalent to training it to map $x_1$ to the clean image $s$, as shown in \cref{eq1}.
\begin{equation}
    \arg\min_{\theta}\mathbb{E}_{x_1,x_2}||f_{\theta}(x_1)-x_2||_2^2=\arg\min_{\theta}\mathbb{E}_{x_1,s}||f_{\theta}(x_1)-s||_2^2
    \label{eq1}
\end{equation}
Therefore, Noise2Noise can train a denoising network that performs comparably to supervised methods trained with noisy-clean image pairs (Noise2Clean).
Unfortunately, in clinical scenarios, even obtaining aligned noisy-noisy image pairs through two scans is often impractical, which poses the greatest obstacle for the clinical application of Noise2Noise.
However, it is worth noting that common medical images, such as those from CT, MRI, and PET, are usually presented in the form of image volumes.
Due to the spatial continuity of tissue structures, there are many regions between neighboring slices of the same image volume that belong to the same tissue, and these regions should theoretically have the same true values.
For convenience, let us refer to the regions between neighboring slices that have the same 2D coordinates and the same true values as matched regions. 
Since there is minimal interference between slices during reconstruction, the corresponding pixels within matched regions exhibit good noise independence.
Specifically, let $\textbf{x}=\{x_i\}_{i=1}^N$ be a noisy image volume consisting of $N$ slices, where $x_i=s_i+n_i$ and $x_{i+1}=s_{i+1}+n_{i+1}$ represent a pair of neighboring slice images in $\textbf{x}$. 
Here, $s_i$ and $s_{i+1}$ are the true values, and $n_i$ and $n_{i+1}$ are zero-mean noise. 
Based on whether the corresponding pixels in $x_i$ and $x_{i+1}$ belong to the same tissue, we can compute a weight matrix $W_i$, where the weight for each pixel in matched regions is 1 and in unmatched regions is 0.
Since $s_i\cdot W_i=s_{i+1}\cdot W_i$, and $n_i\cdot W_i$ and $n_{i+1}\cdot W_i$ are independent, we can apply the Noise2Noise strategy to the matched regions, as shown in \cref{eq2}.
\begin{equation}
    \frac{1}{N-1}\sum_{i=1}^{N-1}||f_{\theta}(x_i)\cdot W_i - x_{i+1} \cdot W_i||_2^2
    \label{eq2}
\end{equation}
Considering the high information redundancy in image data and the findings of previous research\cite{FC22, ZL24}, it is possible to adequately train the denoising network and achieve good performance. 
The key to implementing the above scheme lies in obtaining a reasonable weight matrix $W_i$. In the following section, we explain how to obtain such a weight matrix.

\subsection{Constructing Weighted Training Image Pairs from Neighboring Slice Images}
\label{sec3-2}
\indent \indent For clean image pair $(s_i,s_{i+1})$ from neighboring slices, since $s_i$ and $s_{i+1}$ share the same pixel values in the matched regions, we can calculate the weight matrix $W_i$ using the pixel-wise residual $|s_i-s_{i+1}|^{(u,v)}$. 
Here, $(u,v)$ denotes a position in an image. 
Specifically, we assign a weight of 1 to the pixels where $|s_i-s_{i+1}|^{(u,v)}=0$, and a weight of 0 to the pixels where $|s_i-s_{i+1}|^{(u,v)}>0$.
However, due to noise interference, the residuals of the noisy image pair $(x_i,x_{i+1})$ in the matched regions are not zero, but the residuals of noise.
Therefore, in order to properly distinguish the matched regions from the unmatched regions between $(x_i,x_{i+1})$, it is necessary to avoid the interference caused by noise when calculating the residuals. 

To address this, we adopted the weight matrix computation method shown in \cref{fig1}(a).
We first applied strong low-pass filtering (LPF) to the noisy images $x_i$ and $x_{i+1}$, which involved a non-local means filtering \cite{BA05} followed by a median filtering with a kernel size of 3.
Although the low-pass filtering operation may make the images a little blurry, it does not considerably affect the spatial information of tissue structures in the images. 
Therefore, we can still evaluate the matching status of corresponding pixels in the original noisy image pair $(x_i,x_{i+1})$ by examining the pixel-wise residual $|LPF(x_i)-LPF(x_{i+1})|^{(u,v)}$. 
Additionally, to avoid interference from remaining noise, we did not strictly define regions with $|LPF(x_i)-LPF(x_{i+1})|^{(u,v)}=0$ as matched regions when calculating the weight matrix $W_i$.
Instead, we used a small threshold $th$ to divide the regions, resulting in the weight matrix $W_i$ as shown in \cref{eq3}.
\begin{equation}
    W_i^{(u,v)}=\left\{
        \begin{aligned}
            0, |LPF(x_i) - LPF(x_{i+1})|^{(u,v)} > th \\
            1, |LPF(x_i) - LPF(x_{i+1})|^{(u,v)} \leq th \\
        \end{aligned}
    \right.
    \label{eq3}
\end{equation}
The threshold $th$ is determined by adjusting and observing whether $W_i$ satisfies the following conditions: (1) the regions where $W_i^{(u,v)}=0$ should sufficiently encompass the structural differences between neighboring slices; (2) $W_i$ should avoid misinterpreting differences between the remaining noise after LPF as structural differences between neighboring slices. We provide an ablation study on $th$ in the supplementary materials, which demonstrates that our strategy for setting $th$ is effective and can be easily implemented based on visual inspection alone, and NS-N2N is not sensitive to the choice of $th$.

It is worth noting that LPF in the above process can also be replaced by other traditional or self-supervised denoising methods with better performance.
We tried this approach, but found that it did not greatly improve the denoising network's performance and introduced additional time costs (see supplementary materials for details).
Therefore, using simple low-pass filtering is a cost-effective choice.

\subsection{Self-Supervised Training Scheme with Regularization Constraints}
\indent \indent Given any noisy image volume $\textbf{x}=\{x_i\}_{i=1}^N$, following the procedure described in \cref{sec3-2}, we can obtain weighted training data composed of neighboring slice images $\{(x_i,x_{i+1})\}_{i=1}^{N-1}$ and the weight matrices $\{W_i\}_{i=1}^{N-1}$.
For each pair of neighboring slice images $(x_i,x_{i+1})$, we first apply the Noise2Noise training strategy by minimizing the reconstruction loss $\mathcal{L}_{Recon}$ as shown in \cref{eq4}, to train the denoising network $f_\theta$ to learn the mutual mapping between the matched regions.
\begin{equation}
    \begin{aligned}
        \mathcal{L}_{Recon}=\frac{1}{N-1}\sum_{i=1}^{N-1}&(\frac{1}{2}(||(f_{\theta}(x_i)-x_{i+1})\cdot W_i||_2^2  \\
        & +||(f_{\theta}(x_{i+1})-x_i)\cdot W_i||_2^2))\\
    \end{aligned}
    \label{eq4}
\end{equation}
Since the noise in neighboring slice images is independent and zero-mean, minimizing $\mathcal{L}_{Recon}$ is, in terms of expectation, equivalent to training the denoising network $f_\theta$ to map the noisy signals in the matched regions to the true values\cite{LJ18}. 
Because the neighboring slice images are expected to have the same true values in the matched regions, we introduce the regional consistency loss $\mathcal{L}_{RC}$ as shown in \cref{eq5}, which forces the denoising network to generate identical denoised results for the matched regions, thereby further improving the network's performance and stability.
\begin{equation}
    \mathcal{L}_{RC} = \frac{1}{N-1}\sum_{i=1}^{N-1}||(f_{\theta}(x_i)-f_{\theta}(x_{i+1}))\cdot W_i||_2^2
    \label{eq5}
\end{equation}

Furthermore, the losses $\mathcal{L}_{Recon}$ and $\mathcal{L}_{RC}$ are applied only in the regions where $W_i^{(u,v)}=1$. 
However, the weight matrix $W_i$ cannot perfectly distinguish between matched and unmatched regions, and there may still be a small number of pixel pairs in $(x_i,x_{i+1})$ where the true values differ to some extent, which could be incorrectly identified as matched.
These incorrectly matched pixel pairs may negatively affect the performance of the denoising network $f_\theta$ and lead to incorrect predictions.
To address this, we introduce the inter-slice continuity loss $\mathcal{L}_{IC}$, as shown in \cref{eq6}.
\begin{equation}
    \mathcal{L}_{IC} = \frac{1}{N-1}\sum_{i=1}^{N-1}||f_{\theta}(\frac{x_i+x_{i+1}}{2})-\frac{f_{\theta}(x_i)+f_{\theta}(x_{i+1})}{2}||_2^2
    \label{eq6}
\end{equation}
The inter-slice continuity loss $\mathcal{L}_{IC}$ is derived from first-order approximation.
Specifically, when we perform a Taylor expansion of the denoising function $f_{\theta}(x)$ at $x_i$ and $x_{i+1}$, we obtain: $f_{\theta}(x)=f_{\theta}(x_i)+f_{\theta}^{'}(x_i)(x-x_i)+o((x-x_i))$ and $f_{\theta}(x)=f_{\theta}(x_{i+1})+f_{\theta}^{'}(x_{i+1})(x-x_{i+1})+o((x-x_{i+1}))$.
Since $x_i$ and $x_{i+1}$ are neighboring slices, we can assume that $f_{\theta}^{'}(x_i)\approx f_{\theta}^{'}(x_{i+1})$. Let $x=\frac{x_i+x_{i+1}}{2}$, and adding the two equations above and simplifying, we can derive: $f_{\theta}(\frac{x_i+x_{i+1}}{2})\approx \frac{f_{\theta}(x_i)+f_{\theta}(x_{i+1})}{2}$.
Therefore, $\mathcal{L}_{IC}$ can effectively utilize the prior knowledge of the inherent spatial continuity of tissue structures to reduce the likelihood of erroneous predictions caused by the few incorrectly matched pixel pairs. 
Besides, $\mathcal{L}_{IC}$ is a global regularization term, allowing all regions of neighboring slice images to be involved in the model optimization.
This helps compensate for the limitation of insufficient participation of regions with pronounced structural changes (\ie, high-frequency regions) when only $\mathcal{L}_{Recon}$ and $\mathcal{L}_{RC}$ are used to train the denoising network.

The full objective of the proposed self-supervised training scheme can be written as follows,
\begin{equation}
    \mathcal{L} = \mathcal{L}_{Recon} + \lambda_{RC}\cdot \mathcal{L}_{RC} + \lambda_{IC} \cdot \mathcal{L}_{IC},
    \label{eq7}
\end{equation}
in which $\lambda_{RC}$ and $\lambda_{IC}$ are hyper-parameters that control the strength of $\mathcal{L}_{RC}$ and $\mathcal{L}_{IC}$, respectively.

\section{Experiments}
\indent \indent We validated the performance of NS-N2N on both synthetic and real datasets, comparing its results with state-of-the-art denoising methods. 
Additionally, we conducted ablation studies to analyze the effectiveness of different components, including the proposed weight matrix and various losses.

\subsection{Implementation Details}
\indent \indent The denoising network $f_\theta$ was implemented based on U-Net\cite{RO15}. 
During the training process, we optimized the model using an Adam optimizer with $\beta_1=0.5$ and $\beta_2=0.999$, setting the initial learning rate to $0.001$.
The total number of training epochs was set to $100$, with the learning rate halving every 20 epochs.
The hyper-parameters were set to $\lambda_{RC}=0.5$ and $\lambda_{IC}=1$. 
All experiments were implemented based on the PyTorch framework and conducted on a computer equipped with an Nvidia GeForce RTX 4090.

\begin{figure*}
    \centering
    \includegraphics[width=0.95\linewidth]{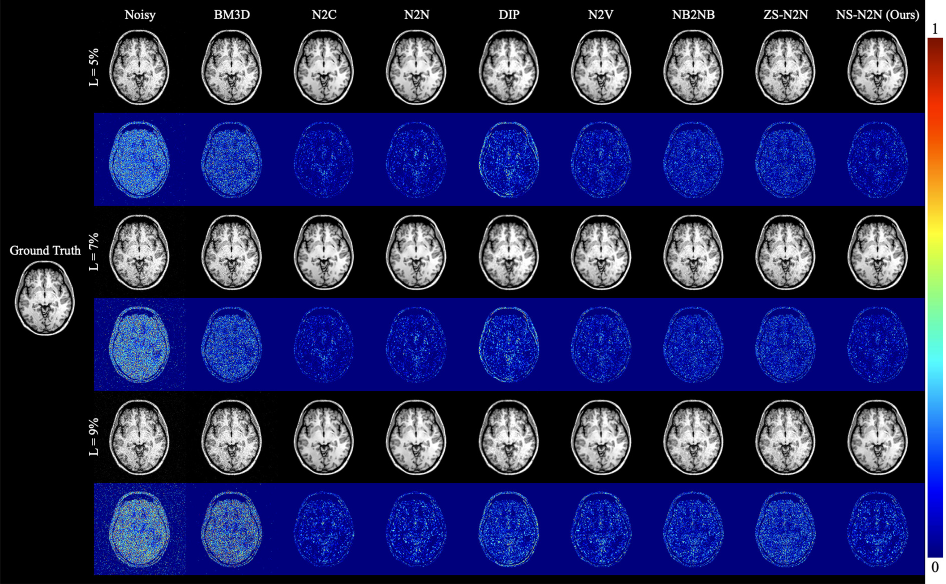}
    \caption{Representative image results of different methods in the synthetic experiment, along with their respective error maps compared to the ground truth.}
    \label{fig2}
    \vspace{-3mm}
\end{figure*}

\begin{table*}
    \centering
    \resizebox{0.7\linewidth}{!}{
        \begin{tabular}{lcccccc}
            \toprule
                &  \multicolumn{2}{c}{L=5\%}  &   \multicolumn{2}{c}{L=7\%}   &     \multicolumn{2}{c}{L=9\%}     \\
            \cmidrule{2-7}
            Method    &   PSNR (dB)   &   SSIM (\%)   &   PSNR (dB)   &   SSIM (\%)   &   PSNR (dB)   &   SSIM (\%)   \\
            \midrule
            Noisy   &   27.88$\pm$1.25  &   84.99$\pm$2.24  &   26.01$\pm$1.09  &   66.42$\pm$1.53  &   24.26$\pm$0.91  &   43.84$\pm$2.41  \\
            BM3D\cite{DK07}    &   29.27$\pm$1.37  &   92.46$\pm$2.37  &   27.11$\pm$1.33  &   87.68$\pm$2.87  &   25.20$\pm$1.15  &   66.63$\pm$1.84  \\
            N2C\cite{RO15}     &   33.68$\pm$2.20  &   96.63$\pm$1.23  &   32.65$\pm$2.16  &   95.97$\pm$1.38  &   32.10$\pm$2.12  &   95.59$\pm$1.53  \\
            N2N\cite{LJ18}     &   32.86$\pm$2.09  &   96.19$\pm$1.34  &   31.79$\pm$2.13  &   95.46$\pm$1.65  &   31.43$\pm$1.93  &   95.18$\pm$1.60  \\
            DIP\cite{UD18}     &   30.91$\pm$1.44  &   95.29$\pm$1.27  &   29.57$\pm$1.38  &   94.06$\pm$1.48  &   28.47$\pm$1.41  &   92.23$\pm$1.76  \\
            N2V\cite{KA19}     &   31.87$\pm$1.69  &   95.65$\pm$1.25  &   31.43$\pm$1.71  &   95.17$\pm$1.43  &   30.60$\pm$1.70  &   94.57$\pm$1.60  \\
            NB2NB\cite{HT21}   &   32.19$\pm$1.39  &   95.55$\pm$1.35  &   31.59$\pm$1.38  &   95.20$\pm$1.35  &   30.74$\pm$1.35  &   94.52$\pm$1.45  \\
            ZS-N2N\cite{MY23}  &   31.27$\pm$1.48  &   94.86$\pm$1.40  &   30.21$\pm$1.52  &   93.87$\pm$1.63  &   29.30$\pm$1.49  &   92.88$\pm$1.82  \\
            NS-N2N (Ours)  &   33.19$\pm$1.64  &   96.60$\pm$1.00  &    32.17$\pm$1.51  &   95.90$\pm$1.11  &   31.46$\pm$1.44  &   95.29$\pm$1.24  \\
            \bottomrule
        \end{tabular}
    }
    \caption{Comparison of PSNR and SSIM for different methods in the synthetic experiment.}
    \label{table1}
    \vspace{-4mm}
\end{table*}

\subsection{Comparison Methods}
\indent \indent We compared the proposed NS-N2N method with two supervised methods (Noise2Clean (N2C)\cite{RO15} and Noise2Noise (N2N)\cite{LJ18}), one traditional method (BM3D\cite{DK07}), and four self-supervised methods that only require single noisy images for training (Deep Image Prior (DIP)\cite{UD18}, Noise2Void (N2V)\cite{KA19}, Neighbour2Neighbour (NB2NB)\cite{HT21}, and Zero-Shot Noise2Noise (ZS-N2N)\cite{MY23}).

\subsection{Synthetic Rician Noise in MRI}
\label{sec4-3}
\vspace{-1mm}
\textbf{Details of Synthetic Experiments}\\
\indent Previous research indicates that noise in MRI follows a Rician distribution\cite{LS20}, where both the real and imaginary components are corrupted by Gaussian noise with the same standard deviation.
Therefore, in this section, we follow\cite{YH22} to simulate noisy images by manually adding Rician noise to T1 modality images from the IXI dataset\cite{IXI18}.
To thoroughly validate the proposed method's performance, we added Rician noise at levels of $L=5\%$, $7\%$, and $9\%$ of the maximum value of the original T1 volumes.
The supervised methods N2C and N2N were trained on noisy-clean and noisy-noisy training sets constructed using 1036 slices from nine image volumes, and were tested on a noisy-clean test set constructed using 220 slices from another image volume.
For the supervised methods, we also considered blind denoising with noise levels ranging from $L\in[5\%,9\%]$ during training.
Since the traditional method (BM3D) and self-supervised methods (NS-N2N, DIP, N2V, NB2NB, ZS-N2N) do not require paired data for training, these methods directly processed the single noisy images in the test set and were tested on the complete noisy-clean test set. 
All images were normalized to $[0,1]$ and resampled to $256\times256$. When calculating the weight matrix $W_i$, for noise levels of $L=5\%$, $7\%$, and $9\%$, we set the threshold $th$ to $0.01$, $0.03$, and $0.05$, respectively.

\begin{figure}
    \centering
    \includegraphics[width=0.8\linewidth]{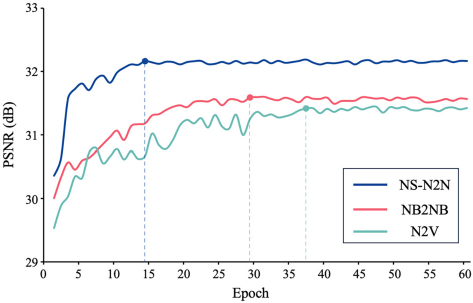}
    \caption{Comparison of training efficiency between the proposed NS-N2N method and two other dataset-based self-supervised methods, NB2NB and N2V, conducted on the synthetic dataset with noise level $L=7\%$.}
    \label{fig3}
    \vspace{-4mm}
\end{figure}

\noindent\textbf{Results of Synthetic Experiments}\\
\indent Representative image results and their error maps with respect to the ground truth are shown in \cref{fig2}, and the quantitative results are listed in \cref{table1}.
The error maps visually demonstrate that, under different noise levels, the proposed NS-N2N method generates denoised results that are more similar to the ground truth compared to traditional and existing self-supervised methods.
From the quantitative results, NS-N2N also achieves higher PSNR and SSIM scores.
This is as expected, because in our method, weighted neighboring slice images can provide the network with abundant training data at the original resolution. In contrast, existing self-supervised methods can only train the network with low-resolution sub-image pairs\cite{HT21,MY23} or a small number of pixel pairs\cite{KA19,BJ19,QY20}.
By comparing the convergence curves shown in \cref{fig3}, we can also observe that the number of training epochs required for the network to converge when using the NS-N2N method is notably fewer than that of the N2V and NB2NB methods.
This indicates that our approach makes more efficient use of the training data and has higher training efficiency than existing dataset-based self-supervised methods.
Compared to the supervised methods, our method performs slightly worse than N2C but better than N2N, which is unexpected.
One possible reason is that our method adequately leverages the spatial continuity information between neighboring slice images, whereas N2N does not adopt such a strategy.

\subsection{Real-world Low-Dose Noise in CT}
\label{sec4-4}
\textbf{Details of Real-world Experiments}\\
\indent In this section, we use the Mayo Low-dose CT Challenge dataset\cite{MC17} to validate the performance of the proposed method in addressing real-world low-dose noise in CT imaging. 
The training set consists of 5326 noisy-clean image pairs from nine image volumes, while the test set includes 610 noisy-clean image pairs from one image volume.
The supervised method N2C was trained on the training set, while the remaining methods directly processed the single noisy images in the test set.
All methods were evaluated on the complete noisy-clean test set.
All images were normalized to the range $[0,1]$ and resampled to $512\times512$.
When calculating the weight matrix, the threshold $th$ was set to $0.01$.

\begin{figure}
    \centering
    \includegraphics[width=\linewidth]{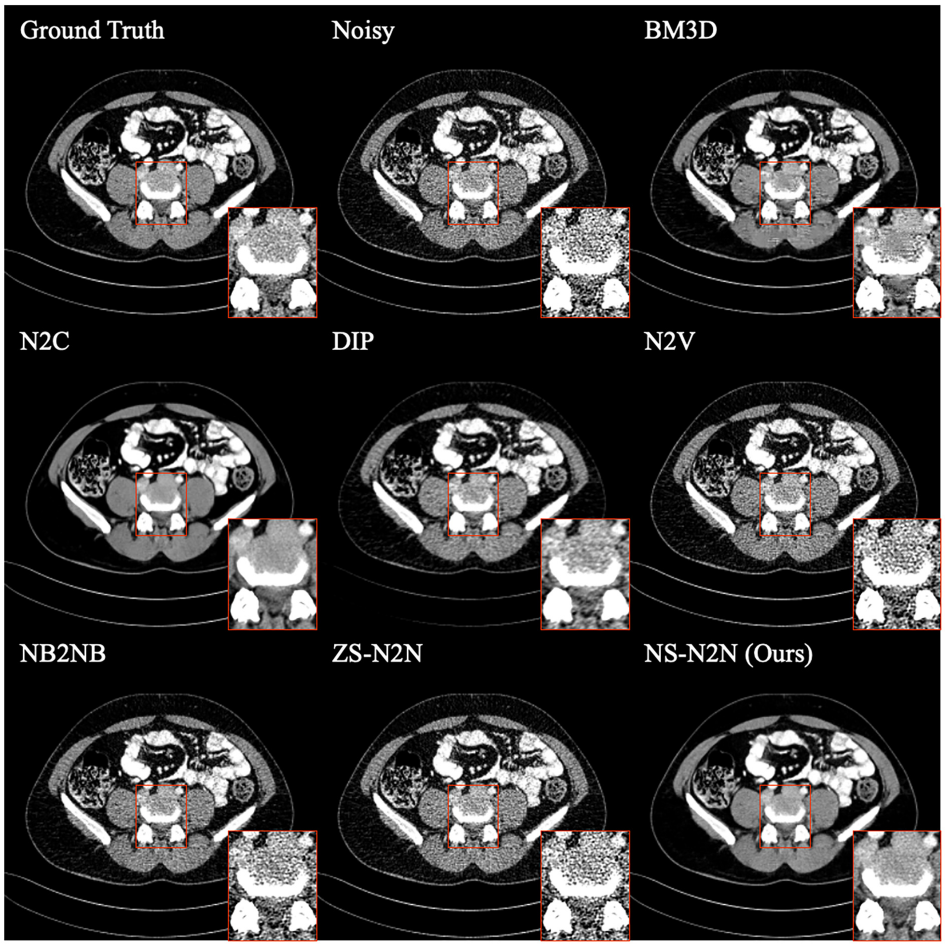}
    \caption{Representative image results of different methods in the real-world experiment.}
    \label{fig4}
    \vspace{-1mm}
\end{figure}

\begin{table}
    \centering
    \resizebox{0.7\linewidth}{!}{
        \begin{tabular}{lcc}
            \toprule
            Method  &   PSNR (dB)   &   SSIM (\%)   \\
            \midrule
            Noisy   &   34.75$\pm$0.95  &   82.11$\pm$3.42  \\
            BM3D\cite{DK07} &   37.25$\pm$0.88  &   89.52$\pm$2.53  \\
            N2C\cite{RO15}  &   40.24$\pm$0.91  &   94.21$\pm$1.27  \\
            DIP\cite{UD18}  &   36.40$\pm$0.74  &   90.64$\pm$1.94  \\
            N2V\cite{KA19}  &   35.83$\pm$0.87  &   86.19$\pm$2.84  \\
            NB2NB\cite{HT21}    &   36.23$\pm$0.88  &   87.33$\pm$2.79  \\
            ZS-N2N\cite{MY23}   &   35.13$\pm$0.89  &   83.30$\pm$3.11  \\
            NS-N2N (Ours)   &   40.06$\pm$0.75  &   93.74$\pm$1.37  \\
            \bottomrule
        \end{tabular}
    }
    \caption{Comparison of PSNR and SSIM for different methods in the real-world experiment.}
    \label{table2}
    \vspace{-4mm}
\end{table}

\noindent\textbf{Results of Real-world Experiments}\\
\indent Representative image results are shown in \cref{fig4}, and the quantitative results are listed in \cref{table2}. It can be observed that when dealing with real-world low-dose noise, the performance of N2V, NB2NB, and ZS-N2N deteriorates considerably. 
This is because the back-projection and filtering operations in CT reconstruction introduce noise correlations between pixels within each slice, which contradicts the pixel-wise independence assumption about noise in these self-supervised methods.
In fact, similar operations that introduce noise correlations also exist in other common medical imaging devices, such as MRI and PET. 
Therefore, these self-supervised methods have limited clinical applicability. In contrast, the proposed NS-N2N method achieved high-quality denoising results.
This can be attributed to the use of neighboring slice images, which offer better noise independence, to train the denoising network. From the quantitative results, only N2C achieved slightly higher PSNR and SSIM scores than our method, this is because it was trained extensively on a large set of paired training data, which is typically unavailable in practical applications.
In addition, the image results of N2C exhibit some over-smoothing, which is consistent with previous research\cite{WD21}, whereas the proposed NS-N2N generates clearer images with better detail preservation. 
Considering that our method only requires a single noisy image volume for training, NS-N2N offers a higher cost-effectiveness and greater clinical applicability than N2C.

\subsection{Comparison of Processing Efficiency}
\begin{figure}[htb]
    \centering
    \includegraphics[width=0.6\linewidth]{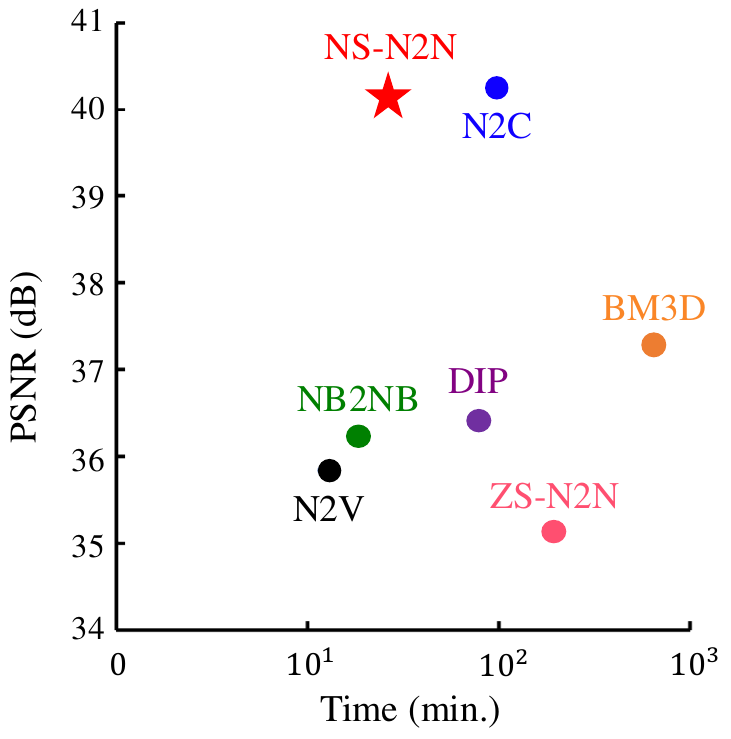}
    \caption{Comparison of total time required for denoising an image volume containing 610 slices using different methods.}
    \label{fig5}
    \vspace{-4mm}
\end{figure}

\indent \indent In addition to denoising performance, processing efficiency is also crucial for clinical applications. In this section, we compare the total time required by different methods to denoise a single noisy image volume, including both training time and testing time, as well as their denoising performance. 
Data used in this section is the test data from \cref{sec4-4}, which is a CT noisy image volume containing 610 slice images.
For N2C, N2V, NB2NB, and NS-N2N, we use the time required for 60 training epochs as their training time, at which point all methods have converged.
The results are visually presented in \cref{fig5}. BM3D, DIP, and ZS-N2N can efficiently denoise a single noisy image.
However, in clinical applications where the entire image volume consisting of hundreds of noisy images needs to be denoised in a single pass, these methods do not offer a processing efficiency advantage, and their denoising performance is not ideal.
The PSNR score of NS-N2N is only slightly lower than that of N2C, but it has a clear superiority in terms of time, as it does not require large amounts of paired data for training.
The time required by NS-N2N is mildly longer than that of N2V and NB2NB in \cref{fig5}.
But considering the NS-N2N method allows the denoising network to converge with far fewer training epochs compared to N2V and NB2NB (as shown in \cref{fig3}), the actual denoising time required by our method should be shorter.
In summary, in real-world medical scenarios, the proposed NS-N2N method demonstrates notable advantages in both denoising performance and processing efficiency compared to existing image-domain denoising methods.

\subsection{Ablation Study}
\indent \indent In this section, we conducted ablation experiments using the synthetic MRI dataset with a noise level of $5\%$ from \cref{sec4-3} and the real-world CT dataset from \cref{sec4-4}, to validate the role of different components, including the proposed weight matrix and various losses. 
The quantitative results are listed in \cref{table3}.
From the first row, we can observe that directly applying the Noise2Noise training scheme to neighboring slice images still yields good performance metrics. 
This indicates that there is a high similarity between neighboring slices, with a large number of matched regions.
With the introduction of the weight matrix $W_i$, both the PSNR and SSIM scores in the two datasets show significant improvements.
This suggests that the inclusion of the weight matrix effectively mitigates the interference of unmatched regions during model optimization.
The metrics in the third and fourth rows of \cref{table3} demonstrating that both $\mathcal{L}_{RC}$ and $\mathcal{L}_{IC}$ effectively enhance the denoising network's performance.
The last row of \cref{table3} achieves the highest PSNR and SSIM scores, indicating that the proposed weight matrix and regularization losses can synergistically improve model performance effectively.
\begin{table}[htb]
    \centering
    \resizebox{\linewidth}{!}{
    \begin{tabular}{@{}c@{\hspace{0.3em}}c@{\hspace{0.3em}}c@{\hspace{0.3em}}cccc@{}}
        \toprule
            & & &   \multicolumn{2}{c}{Rician Noise in MRI}    &  \multicolumn{2}{c}{Low-dose Noise in CT}   \\
        \cmidrule{4-7}
        $W_i$   &   $\mathcal{L}_{RC}$  &   $\mathcal{L}_{IC}$  &   PSNR (dB)   &   SSIM (\%)   &  PSNR (dB)   &   SSIM (\%)    \\
        \midrule
               &    &   &   29.77$\pm$1.55  &   94.23$\pm$1.11  &   38.68$\pm$0.78  &   91.60$\pm$2.17  \\
        \checkmark  &&&     31.09$\pm$1.31  &   95.16$\pm$1.19  &   39.28$\pm$0.81  &   92.31$\pm$2.04  \\
        \checkmark & \checkmark &   &   31.87$\pm$1.51  &   95.65$\pm$1.20  &   39.76$\pm$0.76  &   93.34$\pm$1.79  \\
        \checkmark &    & \checkmark    &   32.07$\pm$1.43  &   95.88$\pm$1.08  &   39.47$\pm$0.83  &   92.79$\pm$1.97  \\
        \checkmark & \checkmark & \checkmark & 33.19$\pm$1.64  &  96.60$\pm$1.00    &  40.06$\pm$0.75   &   93.74$\pm$1.37  \\
        \bottomrule
    \end{tabular}
    }
    \caption{Quantitative results of the ablation study on the synthetic MRI dataset and the real-world CT dataset.}
    \label{table3}
    \vspace{-4mm}
\end{table}

\section{Conclusion}
\indent\indent We proposed Neighboring Slice Noise2Noise (NS-N2N), a novel self-supervised medical image denoising method.
By using neighboring slice images to generate weighted training data and applying a self-supervised scheme with regional consistency loss and inter-slice continuity loss to train the network, our method only requires a single noisy image volume to achieve high-quality denoising of the image volume itself.
Extensive experiments demonstrate that, compared to state-of-the-art self-supervised denoising methods, the proposed NS-N2N method offers notable advantages in both denoising performance and processing efficiency.

\newpage

{
    \small
    \bibliographystyle{ieeenat_fullname}
    \bibliography{main}

\begin{thebibliography}{33}
\providecommand{\natexlab}[1]{#1}
\providecommand{\url}[1]{\texttt{#1}}
\expandafter\ifx\csname urlstyle\endcsname\relax
  \providecommand{\doi}[1]{doi: #1}\else
  \providecommand{\doi}{doi: \begingroup \urlstyle{rm}\Url}\fi

\bibitem[at~Imperial College~London and for the Developing Brain~at King's College~London(2018)]{IXI18}
Biomedical Image Analysis~Group at Imperial College~London and Centre for the Developing Brain~at King's College~London.
\newblock Information extraction from images. https://brain-development.org/ixi-dataset/, 2018.
\newblock Accessed: November 1, 2024.

\bibitem[Batson and Royer(2019)]{BJ19}
Joshua Batson and Loic Royer.
\newblock Noise2self: Blind denoising by self-supervision.
\newblock In \emph{International Conference on Machine Learning}, pages 524--533. PMLR, 2019.

\bibitem[Buades et~al.(2005)Buades, Coll, and Morel]{BA05}
Antoni Buades, Bartomeu Coll, and J-M Morel.
\newblock A non-local algorithm for image denoising.
\newblock In \emph{2005 IEEE computer society conference on computer vision and pattern recognition (CVPR'05)}, pages 60--65. Ieee, 2005.

\bibitem[Chen et~al.(2017)Chen, Zhang, Kalra, Lin, Chen, Liao, Zhou, and Wang]{CH17}
Hu Chen, Yi Zhang, Mannudeep~K Kalra, Feng Lin, Yang Chen, Peixi Liao, Jiliu Zhou, and Ge Wang.
\newblock Low-dose ct with a residual encoder-decoder convolutional neural network.
\newblock \emph{IEEE transactions on medical imaging}, 36\penalty0 (12):\penalty0 2524--2535, 2017.

\bibitem[Dabov et~al.(2007)Dabov, Foi, Katkovnik, and Egiazarian]{DK07}
Kostadin Dabov, Alessandro Foi, Vladimir Katkovnik, and Karen Egiazarian.
\newblock Image denoising by sparse 3-d transform-domain collaborative filtering.
\newblock \emph{IEEE Transactions on image processing}, 16\penalty0 (8):\penalty0 2080--2095, 2007.

\bibitem[Feichtenhofer et~al.(2022)Feichtenhofer, Li, He, et~al.]{FC22}
Christoph Feichtenhofer, Yanghao Li, Kaiming He, et~al.
\newblock Masked autoencoders as spatiotemporal learners.
\newblock \emph{Advances in neural information processing systems}, 35:\penalty0 35946--35958, 2022.

\bibitem[Gong et~al.(2020)Gong, Shan, Teng, Tu, Li, Liang, Wang, and Wang]{GY20}
Yu Gong, Hongming Shan, Yueyang Teng, Ning Tu, Ming Li, Guodong Liang, Ge Wang, and Shanshan Wang.
\newblock Parameter-transferred wasserstein generative adversarial network (pt-wgan) for low-dose pet image denoising.
\newblock \emph{IEEE transactions on radiation and plasma medical sciences}, 5\penalty0 (2):\penalty0 213--223, 2020.

\bibitem[Huang et~al.(2021)Huang, Li, Jia, Lu, and Liu]{HT21}
Tao Huang, Songjiang Li, Xu Jia, Huchuan Lu, and Jianzhuang Liu.
\newblock Neighbor2neighbor: Self-supervised denoising from single noisy images.
\newblock In \emph{Proceedings of the IEEE/CVF conference on computer vision and pattern recognition}, pages 14781--14790, 2021.

\bibitem[Jiang et~al.(2018)Jiang, Dou, Vosters, Xu, Sun, and Tan]{JD18}
Dongsheng Jiang, Weiqiang Dou, Luc Vosters, Xiayu Xu, Yue Sun, and Tao Tan.
\newblock Denoising of 3d magnetic resonance images with multi-channel residual learning of convolutional neural network.
\newblock \emph{Japanese journal of radiology}, 36:\penalty0 566--574, 2018.

\bibitem[Kim et~al.(2018)Kim, Wu, Gong, Dutta, Kim, Son, Kim, El~Fakhri, and Li]{KK18}
Kyungsang Kim, Dufan Wu, Kuang Gong, Joyita Dutta, Jong~Hoon Kim, Young~Don Son, Hang~Keun Kim, Georges El~Fakhri, and Quanzheng Li.
\newblock Penalized pet reconstruction using deep learning prior and local linear fitting.
\newblock \emph{IEEE transactions on medical imaging}, 37\penalty0 (6):\penalty0 1478--1487, 2018.

\bibitem[Krull et~al.(2019)Krull, Buchholz, and Jug]{KA19}
Alexander Krull, Tim-Oliver Buchholz, and Florian Jug.
\newblock Noise2void-learning denoising from single noisy images.
\newblock In \emph{Proceedings of the IEEE/CVF conference on computer vision and pattern recognition}, pages 2129--2137, 2019.

\bibitem[Lehtinen et~al.(2018)Lehtinen, Munkberg, Hasselgren, Laine, Karras, Aittala, and Aila]{LJ18}
Jaakko Lehtinen, Jacob Munkberg, Jon Hasselgren, Samuli Laine, Tero Karras, Miika Aittala, and Timo Aila.
\newblock Noise2noise: Learning image restoration without clean data.
\newblock In \emph{International Conference on Machine Learning}, pages 2965--2974. PMLR, 2018.

\bibitem[Li et~al.(2020)Li, Zhou, Liang, and Liu]{LS20}
Sanqian Li, Jinjie Zhou, Dong Liang, and Qiegen Liu.
\newblock Mri denoising using progressively distribution-based neural network.
\newblock \emph{Magnetic resonance imaging}, 71:\penalty0 55--68, 2020.

\bibitem[Liu et~al.(2024)Liu, Li, Zhao, and Luo]{LJ24}
Jiaming Liu, Guang Li, Qingxian Zhao, and Shouhua Luo.
\newblock Data regularization for streak artifacts removal in self-supervised micro-ct denoising.
\newblock \emph{IEEE Transactions on Radiation and Plasma Medical Sciences}, 2024.

\bibitem[Lu et~al.(2019)Lu, Onofrey, Lu, Shi, Ma, Liu, and Liu]{LW19}
Wenzhuo Lu, John~A Onofrey, Yihuan Lu, Luyao Shi, Tianyu Ma, Yaqiang Liu, and Chi Liu.
\newblock An investigation of quantitative accuracy for deep learning based denoising in oncological pet.
\newblock \emph{Physics in Medicine \& Biology}, 64\penalty0 (16):\penalty0 165019, 2019.

\bibitem[Mansour and Heckel(2023)]{MY23}
Youssef Mansour and Reinhard Heckel.
\newblock Zero-shot noise2noise: Efficient image denoising without any data.
\newblock In \emph{Proceedings of the IEEE/CVF Conference on Computer Vision and Pattern Recognition}, pages 14018--14027, 2023.

\bibitem[Mao et~al.(2016)Mao, Shen, and Yang]{MX16}
Xiaojiao Mao, Chunhua Shen, and Yu-Bin Yang.
\newblock Image restoration using very deep convolutional encoder-decoder networks with symmetric skip connections.
\newblock \emph{Advances in neural information processing systems}, 29, 2016.

\bibitem[McCollough et~al.(2017)McCollough, Bartley, Carter, Chen, Drees, Edwards, Holmes~III, Huang, Khan, Leng, et~al.]{MC17}
Cynthia~H McCollough, Adam~C Bartley, Rickey~E Carter, Baiyu Chen, Tammy~A Drees, Phillip Edwards, David~R Holmes~III, Alice~E Huang, Farhana Khan, Shuai Leng, et~al.
\newblock Low-dose ct for the detection and classification of metastatic liver lesions: results of the 2016 low dose ct grand challenge.
\newblock \emph{Medical physics}, 44\penalty0 (10):\penalty0 e339--e352, 2017.

\bibitem[Quan et~al.(2020)Quan, Chen, Pang, and Ji]{QY20}
Yuhui Quan, Mingqin Chen, Tongyao Pang, and Hui Ji.
\newblock Self2self with dropout: Learning self-supervised denoising from single image.
\newblock In \emph{Proceedings of the IEEE/CVF conference on computer vision and pattern recognition}, pages 1890--1898, 2020.

\bibitem[Ronneberger et~al.(2015)Ronneberger, Fischer, and Brox]{RO15}
Olaf Ronneberger, Philipp Fischer, and Thomas Brox.
\newblock U-net: Convolutional networks for biomedical image segmentation.
\newblock In \emph{Medical image computing and computer-assisted intervention--MICCAI 2015: 18th international conference, Munich, Germany, October 5-9, 2015, proceedings, part III 18}, pages 234--241. Springer, 2015.

\bibitem[Sagheer and George(2020)]{SS20}
Sameera V~Mohd Sagheer and Sudhish~N George.
\newblock A review on medical image denoising algorithms.
\newblock \emph{Biomedical signal processing and control}, 61:\penalty0 102036, 2020.

\bibitem[Sano et~al.(2021)Sano, Nishio, Masuda, and Karasawa]{SA21}
Akira Sano, Teiji Nishio, Takamitsu Masuda, and Kumiko Karasawa.
\newblock Denoising pet images for proton therapy using a residual u-net.
\newblock \emph{Biomedical Physics \& Engineering Express}, 7\penalty0 (2):\penalty0 025014, 2021.

\bibitem[Ulyanov et~al.(2018)Ulyanov, Vedaldi, and Lempitsky]{UD18}
Dmitry Ulyanov, Andrea Vedaldi, and Victor Lempitsky.
\newblock Deep image prior.
\newblock In \emph{Proceedings of the IEEE conference on computer vision and pattern recognition}, pages 9446--9454, 2018.

\bibitem[Wang et~al.(2018)Wang, Yu, Wang, Zu, Lalush, Lin, Wu, Zhou, Shen, and Zhou]{WY18}
Yan Wang, Biting Yu, Lei Wang, Chen Zu, David~S Lalush, Weili Lin, Xi Wu, Jiliu Zhou, Dinggang Shen, and Luping Zhou.
\newblock 3d conditional generative adversarial networks for high-quality pet image estimation at low dose.
\newblock \emph{Neuroimage}, 174:\penalty0 550--562, 2018.

\bibitem[Wu et~al.(2019)Wu, Gong, Kim, Li, and Li]{WD19}
Dufan Wu, Kuang Gong, Kyungsang Kim, Xiang Li, and Quanzheng Li.
\newblock Consensus neural network for medical imaging denoising with only noisy training samples.
\newblock In \emph{International Conference on Medical Image Computing and Computer-Assisted Intervention}, pages 741--749. Springer, 2019.

\bibitem[Wu et~al.(2021)Wu, Kim, and Li]{WD21}
Dufan Wu, Kyungsang Kim, and Quanzheng Li.
\newblock Low-dose ct reconstruction with noise2noise network and testing-time fine-tuning.
\newblock \emph{Medical Physics}, 48\penalty0 (12):\penalty0 7657--7672, 2021.

\bibitem[Xu and Adalsteinsson(2021)]{XJ21}
Junshen Xu and Elfar Adalsteinsson.
\newblock Deformed2self: Self-supervised denoising for dynamic medical imaging.
\newblock In \emph{Medical Image Computing and Computer Assisted Intervention--MICCAI 2021: 24th International Conference, Strasbourg, France, September 27--October 1, 2021, Proceedings, Part II 24}, pages 25--35. Springer, 2021.

\bibitem[Yang et~al.(2022)Yang, Zhang, Han, Zhao, Ren, Sheng, and Zhang]{YH22}
Haibo Yang, Shengjie Zhang, Xiaoyang Han, Botao Zhao, Yan Ren, Yaru Sheng, and Xiao-Yong Zhang.
\newblock Denoising of 3d mr images using a voxel-wise hybrid residual mlp-cnn model to improve small lesion diagnostic confidence.
\newblock In \emph{International Conference on Medical Image Computing and Computer-Assisted Intervention}, pages 292--302. Springer, 2022.

\bibitem[Yang et~al.(2018)Yang, Yan, Zhang, Yu, Shi, Mou, Kalra, Zhang, Sun, and Wang]{YQ18}
Qingsong Yang, Pingkun Yan, Yanbo Zhang, Hengyong Yu, Yongyi Shi, Xuanqin Mou, Mannudeep~K Kalra, Yi Zhang, Ling Sun, and Ge Wang.
\newblock Low-dose ct image denoising using a generative adversarial network with wasserstein distance and perceptual loss.
\newblock \emph{IEEE transactions on medical imaging}, 37\penalty0 (6):\penalty0 1348--1357, 2018.

\bibitem[You et~al.(2019)You, Cao, Lu, Mao, and Wanga]{YX19}
Xuexiao You, Ning Cao, Hao Lu, Minghe Mao, and Wei Wanga.
\newblock Denoising of mr images with rician noise using a wider neural network and noise range division.
\newblock \emph{Magnetic resonance imaging}, 64:\penalty0 154--159, 2019.

\bibitem[Yu et~al.(2023)Yu, Zhang, Zhang, and Zhu]{YJ23}
Jie Yu, Huitao Zhang, Peng Zhang, and Yining Zhu.
\newblock Unsupervised learning-based dual-domain method for low-dose ct denoising.
\newblock \emph{Physics in Medicine \& Biology}, 68\penalty0 (18):\penalty0 185010, 2023.

\bibitem[Zhang et~al.(2017)Zhang, Zuo, Chen, Meng, and Zhang]{ZK17}
Kai Zhang, Wangmeng Zuo, Yunjin Chen, Deyu Meng, and Lei Zhang.
\newblock Beyond a gaussian denoiser: Residual learning of deep cnn for image denoising.
\newblock \emph{IEEE transactions on image processing}, 26\penalty0 (7):\penalty0 3142--3155, 2017.

\bibitem[Zhou and Li(2024)]{ZL24}
Langrui Zhou and Guang Li.
\newblock Reliable multi-modal medical image-to-image translation independent of pixel-wise aligned data.
\newblock \emph{Medical Physics}, 2024.

\end{thebibliography}
}

\end{document}